\documentclass[prd,twocolumn,showpacs,superscriptaddress,nofootinbib]{revtex4}

\usepackage{amsmath}
\usepackage{latexsym}
\usepackage{graphicx}
\usepackage{epstopdf}
\begin{document}


\title{Towards a wave-extraction method for numerical relativity.\\ IV.
Testing the quasi-Kinnersley method in the Bondi--Sachs framework}


\author{Andrea~Nerozzi}
\affiliation{Center for Relativity, University of Texas at
Austin, Austin, Texas 78712, USA}
\affiliation{Institute of Cosmology and Gravitation, University
of Portsmouth, Mercantile House, Portsmouth PO1 2EG, UK}
\author{Marco Bruni}
\affiliation{Institute of Cosmology and Gravitation, University
of Portsmouth, Mercantile House, Portsmouth PO1 2EG, UK}
\affiliation{Dipartimento di Fisica, Universit\`a  degli Studi di Roma  ``Tor Vergata",
via della Ricerca Scientifica 1, 00133 Roma, Italy}
\author{Virginia Re}
\affiliation{
School of Physics and Astronomy, The University of
Birmingham, Edgbaston Birmingham B15 2TT, UK}

\author{Lior~M.~Burko}
\affiliation{Department of Physics, University of Alabama in Huntsville, 
Huntsville, Alabama 35899, USA}


\date{\today}


\begin{abstract}
We present a numerical study of the evolution of a non-linearly disturbed black hole described by the Bondi--Sachs metric, for which the outgoing gravitational waves can readily be found using the news function.
We compare the gravitational wave output obtained with the use of 
the news function in the Bondi--Sachs framework, with that obtained from the Weyl
scalars, where the latter are evaluated in a quasi-Kinnersley tetrad. The latter method has the advantage of being applicable to any formulation of Einstein's equations---including the ADM formulation and its various descendants---in addition to being 
robust. Using the non-linearly disturbed Bondi--Sachs black hole as a test-bed, we show that the two approaches give wave-extraction results which are in very good agreement. 
When wave extraction through the Weyl scalars is done in a non quasi-Kinnersley tetrad, the results are markedly different from those obtained using the news function.
\end{abstract}





\maketitle


\section{Introduction}
\label{sec:introduction}
Gravitational--wave detection has been gaining much interest over the last decades. Much effort has been correspondingly made in modelling possible sources. Interesting examples of sources of 
gravitational waves are, e.g., binary systems of merging black holes, spiralling systems of two neutron stars or coalescing black hole--neutron star binaries. 
Analytical tools able to investigate the dynamics of such sources when the merging takes place do not exist, due to the strong non-linearity of the problem. Numerical simulations are therefore 
invaluable in order to extract information about the gravitational-wave signal emitted. Typically, numerical relativity studies are done in three stages: First, one specifies initial data that correspond to the physical system of interest, and that satisfy certain constraint equations. Next, one 
evolves these initial data numerically, using the evolution equations (with or without enforcing the constraints), and finally, one needs to interpret the results of the simulation and extract the relevant physics thereof. This paper---like its prequels \cite{Beetle04, Nerozzi04,Burko04}---is concerned with this last stage, namely, with the problem of wave extraction. 

In order to interpret the results of numerical relativity simulations---at least within the context of a specific formulation of Einstein's equations---a useful approach is that based on the characteristic initial value problem, originally introduced by Bondi and Sachs \cite{Bondi62,Sachs62}. The characteristic initial value problem has been extensively used in numerical relativity, for spherically symmetric systems 
\cite{Corkill83,Stewart84,Gomez92a,Gomez94b,Clarke94,Clarke94b,Gomez96,Burko97}, 
axisymmetric systems \cite{Isaacson83,Gomez94a,Papadopoulos02} and 3D systems
\cite{Bishop97a,Bishop99,Gomez97a,Gomez97b}. This article is based on the results found in
\cite{Papadopoulos02}---where the characteristic initial value problem, using Bondi coordinates, is used in axisymmetry to study a non-linearly disturbed non-rotating black hole metric. 
The non-linear response of the Schwarzschild black hole to the gravitational
perturbation is embodied in a superposition of angular harmonics propagating outside the source. All the information concerning  the angular harmonics and the energy radiated are easily derived by the evolved quantities. In fact, in this particular case one can identify a {\it news} function \cite{Bondi62}, i.e., a function which embodies the information about the gravitational--radiation energy 
emitted \cite{Bishop97a,dInverno96,dInverno97,Bartnik:1999sg,Zlochower:2003yh,Bishop03,Babiuc:2005pg}.

It is not currently possible to have at hand a quantity such as the news function when using other numerical approaches like, specifically, the widely used 3+1 decomposition of Einstein's equations \cite{Arnowitt62}. 
Indeed, one of the outstanding problems of numerical relativity is that of {\em wave extraction}, i.e., 
the problem of how to extract the outgoing gravitational waves from the results of numerical simulations. 
In currently available methods, various approximations are applied to determine the gravitational-wave emission of isolated  sources. One of the simplest approaches applies the  quadrupole formula (that strictly speaking is valid for weak gravitational fields and slow motions) 
\cite{Thorne80, Schutz86a}. This approach has been used effectively, e.g., in models of stellar collapse 
\cite{Dimmelmeier02b}. More sophisticated approaches use the Moncrief formalism \cite{Moncrief74, Moncrief74b} 
to extract first-order gauge invariant variables from a spacetime which is assumed to be a 
perturbation of a Schwarzschild background at large distances \cite{Abrahams90, Abrahams92a, Alcubierre00b, Alcubierre01a}.
The strength of this approach is that it is gauge invariant, i.e., the information extracted is 
related to the physics of the system, and not to the coordinates used. Specifically, it avoids the pitfall of 
misidentifying gauge degrees of freedom as gravitational waves. 
[E.g., in the Lorenz gauge all degrees of freedom, including the (residual-) gauge ones, travel at the same speed---the speed of light---such that they might be confused with physical waves.] This procedure is usually performed 
under the assumption that the underlying gauge, i.e., the particular choice of the 
spacetime coordinate system, leads to a metric which is asymptotically Minkowski 
in its standard form, which is indeed the case for the most commonly used gauges in simulations of isolated systems. 
The gravitational waveform is determined by integrating metric components over a coordinate sphere at 
some appropriately large distance from the central source, and then subtracting the spherical part of the field 
(which is non-radiative).  However, as we have already mentioned, such techniques are well defined 
when the background metric is assumed to be Schwarzschild, while their application to the more 
generic Kerr background metric can {\em at best} be intended as a very crude approximation. 
In addition, in the typical numerical relativity simulation, one does not usually have information about the mass and spin angular momentum of the eventual quiescent black hole, and no prescription is currently 
known to uniquely separate a background from a perturbation.  
Moreover, once the elimination of the spherical background is performed, what is left does not necessarily 
satisfy the perturbative field equations, and may, in fact, be quite large \cite{seidel01}. 

Another important approach is that of Cauchy--charactestic matching (CCM) schemes \cite{Clarke94}. 
The peculiarity of such schemes is that they use a characteristic Bondi-Sachs approach to study the numerical 
space-time far from the source, where the fields are weak and the probability to form
caustics, which would make the code crash, is limited. In this way, using the notion of the Bondi news 
function, it is possible to extract easily the gravitational wave information. In the strong field
part of the computational domain, instead, a usual Cauchy foliation is used, so that the problem of caustic formation 
is irrelevant. The CCM schemes have been used successfully to simulate cylindrically symmetric vacuum space-times 
\cite{Clarke94b} or to study the Einstein--Klein--Gordon system with spherical symmetry \cite{Gomez96}. 
A fully 3D application of the CCM scheme is, however, still unavailable. 

A final approach worth mentioning here aimed at wave extraction is the one involving the Bel--Robinson vector \cite{Smarr77}, which can be considered a generalization to general relativity of the Poynting vector defined in electromagnetism. However its connection with the radiative degrees of freedom is 
still not entirely clear. 

A novel approach has been suggested recently \cite{Beetle02, Beetle04, Nerozzi04}: one extracts information about the gravitational radiation through quantities  that are gauge and 
background-independent. 
One such quantity is the Beetle--Burko scalar, which is also tetrad independent.    
Specifically, no matter how one chooses to separate the perturbation from the background, 
the Beetle--Burko scalar remains unchanged.  
However, as pointed out in various contexts  \cite{Burko04,Cherubini04,Berti05}, 
the physical meaning of the Beetle--Burko scalar is non trivial. For example, in the stationary 
spacetime of a rotating neutron star its non-zero value is due to the deviation of the quadrupole 
from that of Kerr \cite{Berti05}, while clearly no radiation is present. 
Thus the Beetle--Burko scalar awaits further study.
At any case, the Beetle--Burko scalar---while including information only on the radiative degrees of freedom (when the notion of radiation is defined unambiguously)---describes the latter only partially. To obtain a full description of the radiative degrees of freedom,  it is therefore desirable to consider an approach in which one calculates quantities, 
  whose physical interpretation is more straightforward; however,
the price to pay for these advantages is that it 
      is harder to obtain such quantities uniquely: in fact, one still needs to break the 
spin/boost symmetry in a useful way.

This approach, which is the basis of this paper, is that of using the Weyl 
scalars---which,  under certain assumptions, isolate the radiative degrees 
of freedom from the background and gauge ones 
\cite{Newman62a,Sachs62,Szekeres65}---for wave extraction. 
Teukolsky \cite{Teukolsky73} showed that, choosing a particular 
tetrad, namely the {\it Kinnersley} tetrad \cite{Kinnersley69}, to calculate the Weyl scalars, it is possible to associate the Weyl scalar $\Psi_4$ with the outgoing gravitational radiation for a perturbed Kerr space-time.  Other authors have suggested to use the Weyl scalars for wave extraction, most recently in \cite{Fiske05}, or to explore their relation with metric
perturbations \cite{Lousto05}. However, the Weyl scalars depend on the choice of tetrad. Specifically, performing null rotations on the basis vectors of the null tetrad one can change the values of the Weyl scalars. (Recall, that one of the advantages of the Beetle--Burko scalar is that it is tetrad 
{\em independent}.) In fact, extracting the Weyl scalar $\Psi_4$ is meaningless, unless one also describes how to construct the tetrad to which it corresponds. In most tetrads, the Weyl scalar $\Psi_4$ mixes the information of the outgoing radiation with other information, including gauge degrees of freedom. In our proposal the use of the Weyl scalars is intimately related to the construction of the tetrad in which they are to be calculated---the quasi-Kinnersley tetrad---and therein lies its strength.

With the aim of using $\Psi_4$ to extract information from simulations about the gravitational radiation output, starting with a transverse condition $\Psi_1=\Psi_3=0$, recent work \cite{Beetle04, Nerozzi04, Burko04} has addressed the problem of computing Weyl scalars in a tetrad which will eventually converge to the Kinnersley tetrad. This tetrad has been dubbed the {\it quasi-Kinnersley tetrad}, and work is still in progress in order to uniquely identify it for a general metric. Up to now, it is possible, for a general spacetime, to identify a class of tetrads, namely the quasi-Kinnersley {\em frame}  \cite{Beetle04, Nerozzi04}, with the property that in this frame the radiative degrees of freedom (when and where the notion of radiation is unambiguous) are completely separated from the background ones. However, work is still in progress to identify the quasi-Kinnersley tetrad out of this frame. The difficuly in doing so is related to the following  property of the tetrad members of the quasi-Kinnersley frame: they are all connected through type III (``spin/boost") rotations, and the spin/boost symmetry needs to be broken before the Weyl scalar $\Psi_4$ can be extracted in the quasi-Kinnersley tetrad. The 
quasi-Kinnersley frame is therefore a two-parameter family of tetrads, and the value of $\Psi_4$ (and that of $\Psi_0$) depends on the choice of the tetrad member of the frame. Notably, the Beetle--Burko scalar is invariant under type III rotations. That is, all the tetrad members of the quasi-Kinnersley frame share the same Beetle--Burko scalar. We avoid the spin/boost symmetry breaking difficulty in the present paper by applying an {\em ad hoc} technique to find the quasi-Kinnersley tetrad. This ad hoc technique allows us to obtain the Weyl scalar $\Psi_4$ in the quasi-Kinnersley tetrad, where its interpretation according to the gravitational compass is readily available. 

Specifically, in this paper we use the Bondi--Sachs formalism \cite{Bondi62}, as it turns out that in this special case we can identify a (non-transverse) quasi-Kinnersley tetrad in a simple way, and to compute the Weyl scalars directly. The aim of this work is thus to demonstrate---in the context of this practical numerical example---the applicability, and necessity, of the quasi-Kinnersley tetrad method in using the Weyl scalars as wave extraction tools.

The article is organized as follows: Section \ref{sec:bondi} 
introduces our physical scenario, and Section \ref{sec:weyl} 
describes our Weyl scalars computation. In Section \ref{sec:psinews} we
present the expected result which links the Bondi {\it news function} to $\Psi_4$. Finally results and 
conclusions are presented in Section \ref{sec:results}.


\section{The Bondi problem}
\label{sec:bondi} 

\begin{figure*}
\centering
\includegraphics*[width=13cm, height=8cm]{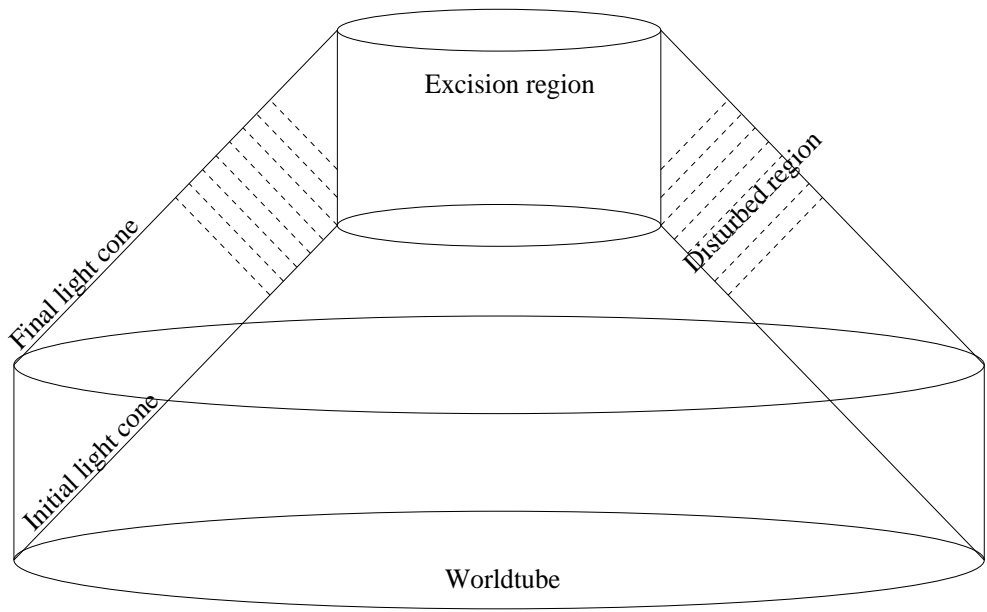}
\caption{Diagram of the Bondi algorithm: the foliation is built up
using ingoing light cones that emanate from an external worldtube $\mathcal{W}$.
We fix the geometry at the world tube to be that of a single 
Schwarzschild black hole. We non-linearly perturb the space-time by setting 
a non-vanishing value for the Bondi function $\gamma$. Such disturbance
propagates outwards in the radiative region, 
after scattering off the black hole.}
\label{fig:diagram}
\end{figure*}

The numerical scenario we are studying is that of a non-linearly perturbed\relax
\footnote{We want to point out here that the expression ``perturbed'' could
 be misleading in this context, as it might suggest we are assuming some kind of approximation.
Our numerical simulations are instead fully non-linear evolutions
of Einstein's equations in the Bondi-Sachs formulation.}
Schwarzschild black hole using an ingoing null-cone foliation of the
space-time. 

We set up our system of coordinates as follows:
a timelike geodesic is the origin of our coordinate system, 
photons are travelling from the origin in all directions, their
trajectories forming null hypersurfaces. The hypersurface 
foliation is labelled by the coordinate $v$. As $r$ coordinate
we choose a luminosity distance, such that the two-surfaces 
of constant $r$ and $u$ have area $4\pi r^2$. Finally, each null
geodesic in the hypersurface is labelled by the two angular 
variables $\theta$ and $\phi$. We will restrict our attention
to an axisymmetric space-time such that $\frac{\partial}{\partial\phi}$
is a Killing vector. Having chosen these variables, the
Bondi metric in ingoing coordinates reads

\begin{eqnarray}\label{metric}
  ds^2&=&-\left[\left(1-2\frac{M}{r}\right)e^{2\beta}-U^2r^2e^{2\gamma}\right]
         \,dv^2\nonumber \\&+&2e^{2\beta}dvdr 
      - 2Ur^2e^{2\gamma}dvd\theta \label{eqn:bondimetric}\\&+&
  r^2(e^{2\gamma}d\theta^2+e^{-2\gamma}\sin\theta^2d\phi^2), \nonumber
\end{eqnarray}
where $M,U,\beta,\gamma$ are unknown functions of the coordinates $\left(v,r,
\theta\right)$.
Within this framework, the 
Einstein equations decompose into three hypersurface equations and one
evolution equation, as given below in symbolic notation

\begin{subequations}
\label{eqn:fieldeqn}
\begin{eqnarray}
\Box^{\left(2\right)}\psi &=& \mathcal{H}_{\gamma}\left(M,\beta,U,\gamma\right), \label{eqn:gammaeqn} \\ 
\beta_{,r} &=& \mathcal{H}_{\beta} \left(\gamma\right), \label{eqn:betaeqn} \\
U_{,rr} &=& \mathcal{H}_U \left(\beta, \gamma\right), \label{eqn:ueqn} \\
M_{,r} &=& \mathcal{H}_M \left(U, \beta, \gamma\right), \label{eqn:meqn}
\end{eqnarray}
\end{subequations}
where $\Box^{\left(2\right)}$ is a 2-dimensional wave operator, $\psi=r\gamma$,
and the various $H$ symbols are functions of the Bondi variables.
We will write here only the expression for $\mathcal{H}_{\beta}$, the simplest
of all the functions, which is given by

\begin{equation}
\mathcal{H}_{\beta}=\frac{1}{2}r\gamma^2.
\label{eqn:hbeta}
\end{equation}
We will use this expression to describe the numerical algorithm.
For exhaustive description of the system (\ref{eqn:fieldeqn})
we refer to \cite{Gomez94a}
and \cite{Isaacson83}.  

The structure of Eq.~(\ref{eqn:fieldeqn}) establishes a natural
hierarchy in integrating them. By setting the initial value for
the function $\gamma$ on the initial hypersurface 
(in addition to four free parameters), it is possible to
integrate Eq.~(\ref{eqn:betaeqn}) to obtain $\beta$, then, having
both $\beta$ and $\gamma$, Eq.~(\ref{eqn:ueqn}) can be integrated to obtain
$U$ and finally $M$ can be derived by integrating Eq.~(\ref{eqn:meqn}). At
this point we have all the metric functions on the initial hypersurface
and we can integrate Eq.~(\ref{eqn:gammaeqn}) to obtain $\gamma$ on the
next hypersurface, and the procedure is iterated.
The integration of Eq.~(\ref{eqn:fieldeqn}) introduces constants of
integration, which we set to zero (Bondi frame \cite{Tamburino66,Isaacson83}),
which corresponds to having an asymptotically inertial frame.

The metric introduced in Eq.~(\ref{eqn:bondimetric}) 
describes a static Schwarzschild black hole if 
we set, in the Bondi frame,  all the functions except $M$ 
to zero everywhere in the domain.
$M$ is chosen to be the Schwarzschild mass $M_0$ of the black hole.
Besides, outgoing gravitational radiation as a perturbation is introduced
by the function $\gamma$; it turns out that $\gamma$ is a spin-2 field 
and is actually related to the radiative degree of freedom. Choosing
an initial shape for $\gamma$ means in practice choosing the initial profile of
outgoing gravitational waves.
More specifically, the initial data are chosen in the following way: 

\begin{itemize}
\item We recover the background Schwarzschild geometry by setting $\gamma=\beta=
U=0$ over the whole computational domain, while we set $M=M_0$
\item We set up the initial data for a gravitational wave outgoing pulse 
by choosing  
a gaussian shape (with parameters $r_c$ and $\sigma$) for the function $\gamma$
(the choice of the gaussian shape is dictated by requiring the function to vanish
at the outer boundary of the grid):

\begin{equation}
\gamma\left(r,\theta\right) = \frac{\lambda}{\sqrt{2\pi}\sigma}e^{-\frac{\left(r
-r_c\right)^2}{\sigma^2}} Y_{2lm}\left(\theta\right),
\label{eqn:gammapert}
\end{equation}
where $\lambda$ is the amplitude of the perturbation, and $Y_{lm}$ is the
spherical harmonic of spin 2. In our numerical simulations we will set
two types of initial data: the first one with $l=2$
and $m=0$ to get a pure quadrupole outgoing perturbation, while in the
second one we will set $l=3$ and $m=0$.
\end{itemize}

The integration of the hypersurface equations leads to the problem
of the gauge freedom in choosing the integration constants. This
apparent freedom is fixed by choosing outer boundary conditions on
our numerical grid. As depicted in Fig.~(\ref{fig:diagram}) 
we in fact fix the metric at the outer
boundary, i.e. on the worldtube $\mathcal{W}$,
to be that of a Schwarzschild black hole. This automatically fixes
the integration constants to be 0 for $\gamma$, $U$ and $\beta$, and
$M_0$ for $M$ (Bondi frame). We point out here 
that such a boundary condition is well
posed only for simulations which are limited in time, so that no
relevant outgoing gravitational flow has crossed the worldtube.
More information about the evolution routine can be found in
\cite{Papadopoulos94, Papadopoulos02, NerozziTh04}.


\section{Weyl scalars}
\label{sec:weyl}

Once we have
the numerically computed metric for the evolved space-time we
can derive the Newman-Penrose
quantities we need. 
The Weyl scalars are 

\begin{subequations}
\label{eqn:weylscalars}
\begin{eqnarray}
\Psi_0 &=& -C_{abcd}l^am^bl^cm^d, \label{eqn:psi0} \\
\Psi_1 &=& -C_{abcd}l^an^bl^cm^d, \label{eqn:psi1} \\
\Psi_2 &=& -C_{abcd}l^am^b\bar{m}^cn^d, \label{eqn:psi2}\\
\Psi_3 &=& -C_{abcd}l^an^b\bar{m}^cn^d, \label{eqn:psi3} \\
\Psi_4 &=& -C_{abcd}n^a\bar{m}^bn^c\bar{m}^d, \label{eqn:psi4} 
\end{eqnarray}
\end{subequations}
where $l^a$, $n^a$, $m^a$ and $\bar{m}^a$ are the Newman-Penrose  
null vectors.
The five
scalars defined in Eq.~(\ref{eqn:weylscalars})
are of course coordinate independent,
but they do depend on the particular tetrad choice. 

We calculate the scalars in a quasi-Kinnersley tetrad, i.e., 
in a tetrad that converges to the Kinnersley tetrad when 
space-time settles down to that of an unperturbed black hole. In 
\cite{Beetle04, Nerozzi04} a procedure to find the quasi-Kinnersley 
frame in a background independent way is given by looking at transverse frames, 
i.e., those frames where $\Psi_1$ and $\Psi_3$ vanish. In the following we use the
word {\it frame}  to indicate an equivalence class of tetrads which are connected
by a type III spin/boost tetrad transformation. Fixing the right
quasi-Kinnersley tetrad means choosing the tetrad in the
quasi-Kinnersley frame which shows the right radial behavior 
for $\Psi_0$ and $\Psi_4$ according to the peeling-off theorem.  
That is, we seek the tetrad in which these two Weyl scalars peel off correctly.  
A background independent procedure to single out the
tetrad out of the frame currently needs further investigation.

In the Bondi--Sachs framework
the identification of a quasi-Kinnersley tetrad is simple, and
does not need to use the notion of transverse frames.
The main reason for this proprety is due to the asymptotic knowledge of the
Bondi functions when spacetime approaches Schwarzschild: in fact
$\gamma$, $U$ and $\beta$ tend to zero, while $M$ tends to the
Schwarzschild mass $M_0$ of the black hole (for further details see
the next section).  

This situation is much different from the typical 
situation in numerical relativity simulations, for which  the wave-extraction 
methods needs to be background-independent. 
If we assume to be in the Schwarzschild limit, 
the background Kinnersley tetrad chosen by Teukolsky \cite{Teukolsky73}
in the perturbative scenario
would look, using our coordinates $\left(v,r,\theta,\phi\right)$, as

\begin{subequations}
\label{eqn:schwarztetradnull}
\begin{eqnarray}
\ell^{\mu}&=&\left[\frac{2r}{r-2M},1,0,0\right],
\label{eqn:schwarztetradnulll} \\
n^{\mu}&=&\left[0,-\frac{r-2M}{2r},0,0\right],
\label{eqn:schwarztetradnulln} \\
m^{\mu}&=&\left[0,0,\frac{1}{\sqrt{2}r},\frac{i}
{\sqrt{2}r\sin\theta}\right].
\label{eqn:schwarztetradnullm}
\end{eqnarray}
\end{subequations}
This tetrad has been chosen by letting the $\ell^{\mu}$ and $n^{\mu}$
vector coincide with the repeated principal null directions of the
Schwarzschild space-time. Such a condition fixes a frame, i.e. a 
set of tetrads connected by a type III rotation. The type III rotation
parameter is then fixed by setting the spin coefficient $\epsilon$ to
be vanishing.
Eq.~(\ref{eqn:schwarztetradnull}) can be used to find 
the general expression for
the tetrad in the full Bondi formalism. Using the asymptotic values
of the Bondi functions, we can write down the expression for a general
tetrad for the Bondi metric, whose vectors converge to the null
vectors written in Eq.~(\ref{eqn:schwarztetradnull}) in the
Schwarzschild limit. The result
is given by

\begin{subequations}
\label{eqn:bonditetradnull}
\begin{eqnarray}
\ell^{\mu}&=&\left[\frac{2}{\left[\left(1-2M/r\right)e^{4\beta}-
U^2r^2e^{2(\gamma+\beta)}\right]},e^{-4\beta},0,0\right], \nonumber \\
\label{eqn:bonditetradnulll} \\
n^{\mu}&=&\left[0,-\frac{\left[\left(1-2M/r\right)e^{2\beta}-
U^2r^2e^{2\gamma}\right]}{2},0,0\right],
\label{eqn:bonditetradnulln} \\
m^{\mu}&=&\left[0,\frac{rUe^{(\gamma-2\beta)}}
{\sqrt{2}},\frac{1}{\sqrt{2}re^{\gamma}},\frac{i}
{\sqrt{2}r\sin\theta e^{-\gamma}}\right].
\label{eqn:bonditetradnullm}
\end{eqnarray}
\end{subequations}
Henceforth we will denote this first tetrad choice, which is 
supposed to be the successful one, as the tetrad $\mathcal{T}_1$.
It is worth pointing out that the tetrad $\mathcal{T}_1$ is not
transverse, i.e. $\Psi_1$ and $\Psi_3$ are not vanishing; Nevertheless it satisfies the requirements needed of a quasi-Kinnersley tetrad. (The quasi-Kinnersley tetrad does not have to be transverse, although it does need to be {\em asymptotically} transverse.)

As explained above, the choice of $\mathcal{T}_1$
has been driven by the form of the Kinnersley tetrad expressed in
Eq.~(\ref{eqn:schwarztetradnullm}) for the unperturbed black hole. 
However, this is not
always possible in general where tetrad choices are dictated 
by different criteria. A straightforward example in this case
would be that of basing the tetrad on null vectors directly
derived from the metric. For instance,  in the specific example of the metric (\ref{metric}), an apparently natural choice of the tetrad, obtained
with the algebraic manipulation packages  Maple and GRTensor, is
the following:

\begin{subequations}
\label{eqn:bonditetradnull2}
\begin{eqnarray}
\ell^{\mu}&=&\left[
0,-e^{-4\beta},0,0\right], \nonumber \\
\label{eqn:bonditetradnulll2} \\
n^{\mu}&=&\left[e^{2\beta},\frac{\left[\left(1-2M/r\right)e^{2\beta}-
U^2r^2e^{2\gamma}\right]}{2},0,0\right], \nonumber \\
\label{eqn:bonditetradnulln2} \\
m^{\mu}&=&\left[0,
\frac{rUe^{(\gamma-2\beta)}}{\sqrt{2}},
\frac{1}{\sqrt{2}re^{\gamma}},
\frac{i}{\sqrt{2}r\sin\theta e^{-\gamma}}\right].
\label{eqn:bonditetradnullm2}
\end{eqnarray}
\end{subequations}
The reason why this tetrad choice {\it looks} more natural than the first one
is related to the fact that packages like GRTensor construct this tetrad
starting from the $\ell^{\mu}$ vector, which is assumed to be lying
on the null foliation, leading to the expression $\ell_{\mu}=\delta_{\mu 0}$.
The contravariant components are then given by 
Eq.~(\ref{eqn:bonditetradnulll2}). Once $\ell^{\mu}$ is fixed, the other
tetrad vector expressions are found by imposing the normalization conditions
between the vectors in the Newman-Penrose formalism.

The tetrad in Eq.~(\ref{eqn:bonditetradnull2}) will be 
hereafter referred to as tetrad $\mathcal{T}_2$. 
We will show that this different tetrad choice
 leads to results which, although equivalent from
a qualitative point of view, are different than those obtained in the quasi-Kinnersley tetrad $\mathcal{T}_1$. 

In the numerical results that we are going to present in the
next sections we have calculated the Weyl scalars in the
two tetrads presented above, to show the reliability of quasi-Kinnersly tetrad method and an example, through a  complete comparison, of the different results that we might obtain in a numerical simulation in doing wave extraction using Weyl scalars in different tetrads.


\section{The linear regime}
\label{sec:psinews}

In the linear regime both the Bondi and
the Newman--Penrose formalisms  define quantities which provide
 information about gravitational waves. We will discuss here
briefly such definitions in order to get a correspondence between
the two approaches, which will be tested numerically in the 
following section.

In the Bondi formalism the initial assumption is that the space-time
is asymtotically flat, which leads to the following expansion
for the function $\gamma$ at null infinity:

\begin{equation}
\gamma=K+\frac{c}{r}+O\left(r^{-2}\right).
\label{eqn:gammaradial}
\end{equation}
Here we assume to be in the Bondi frame \cite{Bondi62}, 
i.e. we set the integration constant $K$ to be zero.
By integrating the hypersurface equations for the other Bondi functions
we can get their radial expansion at null infinity. Such integrations
in general introduce other integration constants but, again, we assume that
in our frame those constants are vanishing, ending up with the following
expressions for the remaining Bondi functions \cite{Isaacson83}: 

\begin{subequations}
\label{eqn:bondiradialexp}
\begin{eqnarray}
\beta&=&-\frac{c^2}{4r^2}+O\left(r^{-4}\right), \label{eqn:betaexp} \\
U&=&
-\frac{\left[c\sin^2\theta\right]_{,\theta}}{r^2\sin^2\theta}+O\left(r^{-3}\right), \label{eqn:Ueqrad} \\
M&=&M_0+O\left(r^{-1}\right). \label{eqn:Veqrad}
\end{eqnarray}
\end{subequations}
It is trivial to verify that, with this choice of integration 
constants, the space-time is asymptotically flat.
It is possible to define at null infinity a notion of energy, which
leads to the result found by Bondi 

\begin{equation}
E=\frac{1}{4\pi}\oint M\sin\theta d\theta d\phi,
\label{eqn:bondimassfunc}
\end{equation}
and in addition the energy flux per unit solid
angle, which is given
by

\begin{equation}
\frac{d^2E}{dvd\Omega}=-\frac{\left(c_{,v}\right)^2}{4\pi}.
\label{eqn:bondienfunc}
\end{equation}

It is clear that the information about the energy carried by
gravitational waves is contained in the Bondi news function $c_{,v}\approx r\gamma_{,v}$,
where the approximation is assumed to hold at large distances in the
linear regime. Expressing Eq.~(\ref{eqn:bondienfunc}) in terms of
$\gamma$ gives

\begin{equation}
\frac{d^2E}{dvd\Omega}\approx-\frac{r^2\left(\gamma_{,v}\right)^2}{4\pi}.
\label{eqn:bondienfuncgamma}
\end{equation}

An analogous derivation can be achieved within the Newman-Penrose
formalism. The key point is that 
$\Psi_4$ can be expressed, when computed 
in the {\it quasi-Kinnersley} tetrad, directly as a function
of Riemann tensor components, i.e. \cite{Sachs62,Teukolsky73}

\begin{equation} 
\left(\Psi_4\right)_{qKT}=-\left(R_{\hat{v}\hat{\theta}\hat{v}\hat{\theta}}-
iR_{\hat{v}\hat{\theta}\hat{v}\hat{\phi}}\right).
\label{eqn:psi4quasi}
\end{equation} 
The hatted symbols in Eq.~(\ref{eqn:psi4quasi}) are indicating that
the Riemann tensor components are contracted over a tetrad of vectors
oriented along the coordinates. In the linear regime however, those
vectors can be assumed to be the basis coordinate vectors, as the
perturbation is already contained in the Riemann tensor. For this
reason we will omit the hatted symbols from now on, and always talk
about coordinate components.

The components of the Riemann tensor in Eq.~(\ref{eqn:psi4quasi}) 
can then be related
to the transverse-traceless gauge terms, using
$R_{v\alpha v\beta}=-\frac{1}{2}\frac{\partial^2 h_{\alpha\beta}}{\partial v^2}$, which leads to the result 

\begin{equation}
\left(\Psi_4\right)_{qKT}=-\frac{1}{2}\left(\frac{\partial^2 h_{\theta\theta}}
{\partial v^2}-i\frac{\partial^2 h_{\theta\phi}}
{\partial v^2}\right).
\label{eqn:psi4hmunu}
\end{equation}
This relation between $\Psi_4$ in the {\it quasi-Kinnersley} tetrad
and the transverse--traceless (TT) components of the perturbed metric 
leads us to a definition
of the energy emitted by simply calculating the expression of the
energy tensor for the gravitational wave defined by

\begin{equation}
T^{GW}_{\mu\nu}=\frac{1}{32\pi}\left[\partial_{\mu}\left(h^{TT}\right)^{\sigma\rho}
\partial_{\nu}\left(h^{TT}\right)_{\sigma\rho}\right].
\label{eqn:stressenergygrav}
\end{equation}

The total energy flux is then given by the
formula, which is assumed to hold at null infinity:
                                                                                
\begin{equation}
\frac{d^2E}{dvd\Omega}=-
r^2{\left(T^{GW}\right)^r}_v=\frac{r^2}{16\pi}\left[\left(\frac{\partial h^{TT}_{\theta\theta}}{\partial v}\right)^2
+\left(\frac{\partial h^{TT}_{\theta\phi}}{\partial v}\right)^2\right],
\label{eqn:energyfluxv}
\end{equation}
and, by substituting our expression in terms of $\Psi_4$ we get
the result

\begin{equation}
\frac{d^2E}{dvd\Omega}=-
\frac{r^2}{4\pi}\left|\int^v_0\left(\Psi_4\right)_{qKT}dv\right|^2.
\label{eqn:energypsi4}
\end{equation}
In our specific case of axisymmetry we expect only one polarization
state to be present, which is made evident by the presence of a 
single news function. Correspondingly, we expect $\left(\Psi_4\right)_{qKT}$
to have only its real part non vanishing.
Combining Eq.~(\ref{eqn:energypsi4}) with Eq.~(\ref{eqn:bondienfuncgamma})
and considering the presence of only one polarization state, we
finally obtain that in the linearized regime the
relation

\begin{equation}
\left(\Psi_4\right)_{qKT}=-\frac{\partial^2\gamma}{\partial v^2}
\label{eqn:psi4gamma}
\end{equation}
must hold. This is the relation we want to verify numerically. The minus sign comes from the negative sign given in
Eq.~(\ref{eqn:psi4hmunu}). We want to stress again the attention to the
fact that such relation is strictly true at null infinity, however, we
expect it to be well satisfied provided we are at sufficiently large
distances from the black hole. As it will be clear in the next section,
this assumption turns out to be very well motivated.


\begin{figure}
\centering
\includegraphics*[width=8cm, height=8cm]{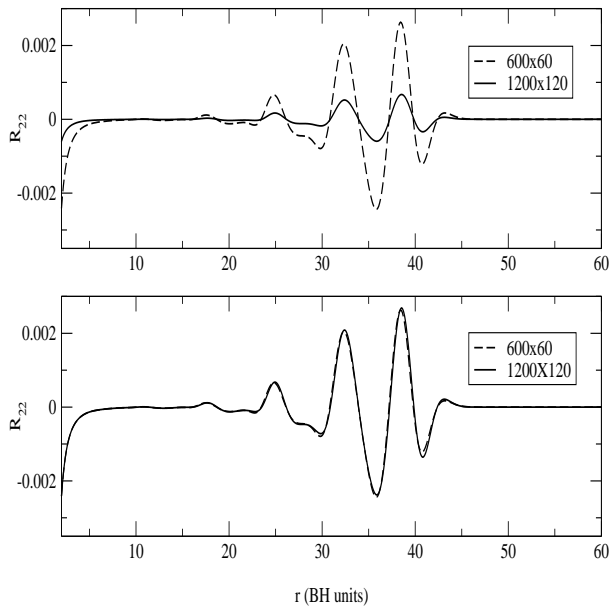}
\caption{Convergence test for the $R_{22}$ component of the
Ricci tensor. The top panel shows this component for two different
resolutions for a radial slice on the equatorial plane. 
As expected the value is converging to zero. In the
bottom panel we have tested the second order power law of convergence
by multiplying the $1200\text{x}120$ output by a factor of four. The two
curves now overlap perfectly, thus proving second order convergence.}
\label{fig:ricciconv}
\end{figure}

\begin{figure}
\centering
\includegraphics*[width=8cm, height=8cm]{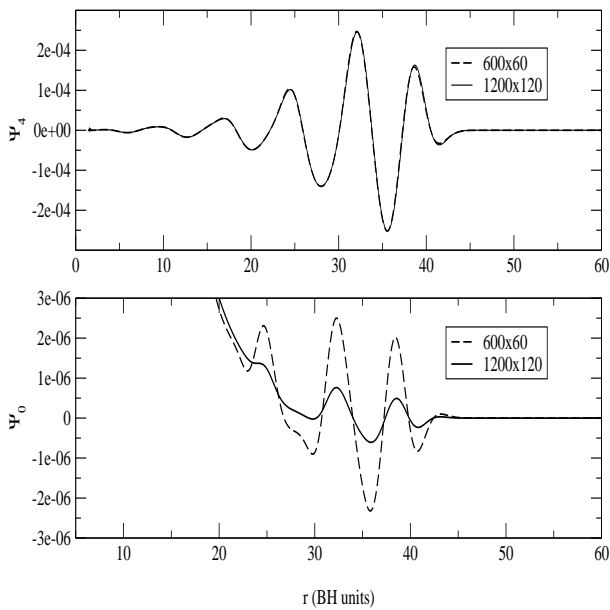}
\caption{(a) Top panel: the value for $\Psi_4$ for two different 
resolutions at
time $v=80$. (b) Bottom panel: the value for $\Psi_0$ for two 
different resolutions at
$v=80$. Both values are calculated on the equatorial plane.} 
\label{fig:convpsi04}
\end{figure}

\section{Numerical Results}
\label{sec:results}

In this section we present numerical results
for a standard simulation. We have written a code that
solves the Bondi equations and calculates the Weyl scalars
in the two tetrads $\mathcal{T}_1$ and $\mathcal{T}_2$.
Our code makes use of the Cactus infrastructure \cite{cactusweb}.

We set up an initial Schwarzschild black hole, and construct an
initial quadrupole perturbation on $\gamma$ using the expression
indicated in Eq.~(\ref{eqn:gammapert}). The values chosen
in this case are $\lambda=0.1$, $r_0=3$ and $\sigma=1$, although
various tries have been performed varying these parameters, all 
leading to the same physical results. 
We emphasize that $\lambda$ represents the amplitude of the initial perturbation: the chosen value is such that the perturbation is somehow realistically small, yet large enough for the full non-linearity of the problem to appear clearly through the harmonic coupling, as we are going to show (see Figg.\ \ref{fig:energyl2T1} and \ref{fig:energyl3T1}, cf.\  also \cite{Papadopoulos02}).
All the results presented
here are obtained using two different resolutions, the coarser one
having 600 points in the radial dimension and 60 points in the
angular direction, the finer one having those values doubled.
The results which are not convergence test results are all 
obtained using the finer resolution of 1200 points in the radial
direction and 120 points in the angular direction.

We will first present
some tests in order to verify the robustness of our algorithm,
and then we will proceed to a full comparison of our results
in the two approaches presented here. The first two following subsections
will deal with the calculation of the Weyl scalars in the tetrad
$\mathcal{T}_1$ defined in Eq.~(\ref{eqn:bonditetradnull}); 
we don't expect the second
tetrad to give different results for what concerns radial fall-offs
and convergence. Section \ref{sec:psi4news} will instead
deal with the relation of $\Psi_4$ with the news function and, within
this context, it is very important to show a comparison of results
in different tetrads, to have an evident demonstration of how 
important the choice of the right tetrad is, i.e.\ the quasi-Kinnersley tetrad,  in the process of
evaluating the outgoing gravitational wave contribution.

\subsection{Convergence}
\label{sec:convergence}

The first thing we want to test in our code is of course convergence.
In order to do so, once the numerical variables are computed, we 
have calculated independently the values of the Ricci tensor components,
which should vanish in vacuum.
Such components are suitable for doing convergence tests. In
Fig.~(\ref{fig:ricciconv}) we show the value of the Ricci component
$R_{22}$ for two resolutions, the picture shows a radial slice
of our space-time on the equatorial plane, 
for the time value $v=80$.  The first
figure simply superimposes the two values obtained for the two
different resolutions, while in the second picture we have first 
multiplied the values for the finer resolution by a factor of four,
as expected in a second order convergence code.

We have found similar results for the
other components, which ensured us of the convergence of our
algorithm.

\subsection{Radial fall-offs of the scalars}
\label{sec:radialfalloffs}

Figg.~(\ref{fig:convpsi04}a) and (\ref{fig:convpsi04}b) show the
numerical output for $\Psi_0$ and $\Psi_4$ for two different 
numerical resolutions. The outputs show satisfactory
convergence for $\Psi_4$, but not for $\Psi_0$. This is because the
asymptotic radial behavior for $\Psi_0$ should be $r^{-5}$, as expected from the peeling-off conjecture (see e.g.\ \cite{Wald84}) and the linear perturbation analysis  
\cite{Teukolsky73}, and
this gets completely embedded in the numerical error. 
We believe that this could constitute a serious numerical
problem in situations where the initial tetrad chosen
for the scalars computation is not the right one, and a tetrad
rotation is needed. The numerical error found in $\Psi_0$
would then propagate when other quantities, like the
curvature invariants $I$ and $J$, are computed, thus leading
to meaningless results. Recall, however, that the curvature invariants 
$I,J$, in addition to the Coulomb scalar $\chi$ and the Beetle--Burko 
scalar $\xi$, can be found invariantly and in a background-independent 
way which is also tetrad-independent, i.e., it does not require finding first 
the Weyl scalars to find $I,J$ \cite{Beetle04, Burko04}. 

\begin{figure}
\centering
\includegraphics*[width=8cm, height=8cm]{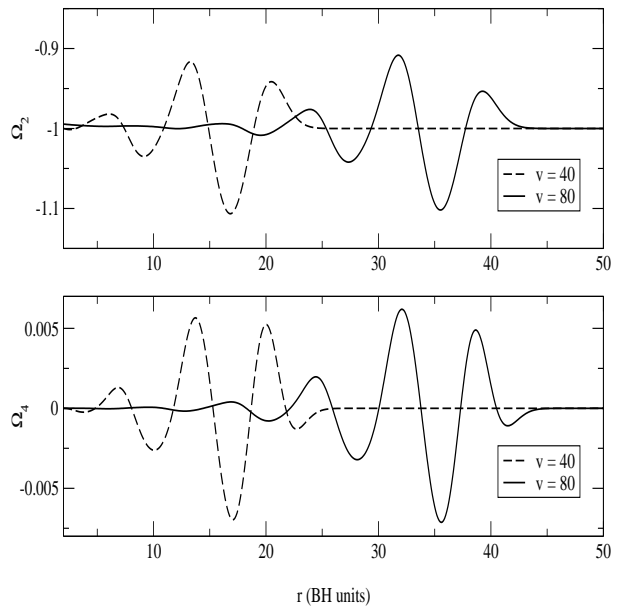}
\caption{
(a) Top panel: the value for $\Omega_2=r^3\Psi_2$ at two different times
$v_1=40$ and $v_2=80$. (b) Bottom panel: the value for 
$\Omega_4=r\Psi_4$ for the
same couple of times $v_1=40$ and $v_2=80$.}
\label{fig:radpsi04}
\end{figure}

Fig.~(\ref{fig:radpsi04}a) and (\ref{fig:radpsi04}b)
emphasize the radial dependence of $\Psi_2$ and $\Psi_4$, which is 
highlighted very well in our simulations. The two figures show
that at late times $\Psi_2$ gets  the background contribution
with the superposition of a wave whose radial behaviour is $r^{-3}$.
We have tested the convergence of such a wave to prove its
physical meaning; this is itself a quite interesting result
as we don't have a perturbation equation for $\Psi_2$, and it is entirely due to the full non-linear treatment of the problem.  Of course,
given its rapid fall-off, the wave contribution from $\Psi_2$ is
negligible.
$\Psi_4$ shows instead the well expected $r^{-1}$ behaviour.

\subsection{Relation of $\Psi_4$ with the Bondi news}
\label{sec:psi4news}

\begin{figure}
\centering
\includegraphics*[width=8cm,height=9cm]{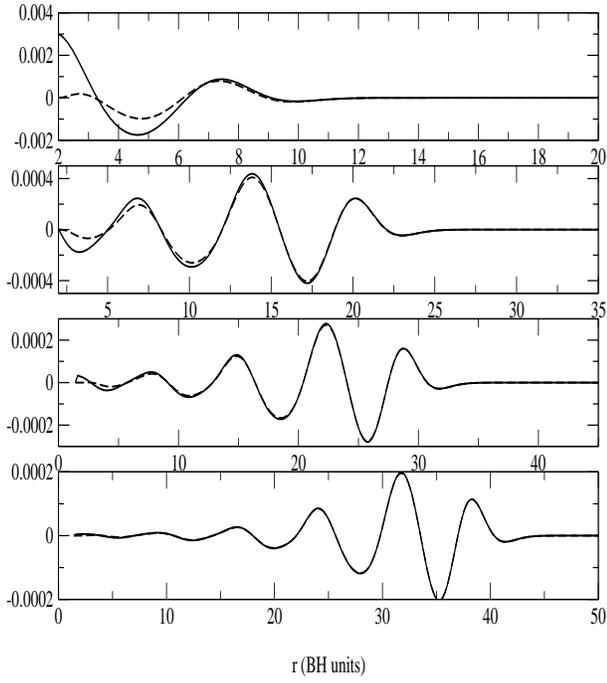}
\caption{From top to bottom: the comparison of $\Psi_4$ (dashed line)
calculated in tetrad
$\mathcal{T}_1$ and $-\gamma_{,vv}$ (solid line) for
the values of $v_0 = 10$, $v_1=40$, $v_2=60$ and $v_3=80$.}
\label{fig:newsbondi}
\end{figure}

\begin{figure}
\centering
\includegraphics*[width=80mm,height=40mm]{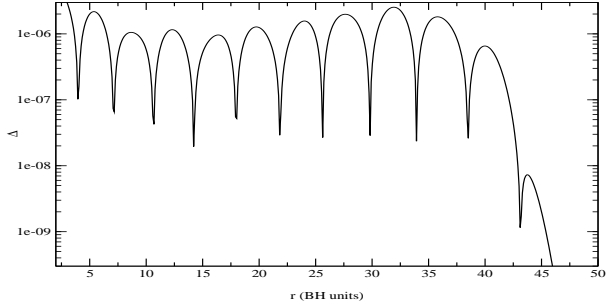}
\caption{The function $\Delta=\left|\Psi_4+\gamma_{,vv}\right|$ at
$v_3=80$.}
\label{fig:newsdiff}
\end{figure}

In this section we want to show the comparison of $\Psi_4$ with the second time derivative of $\gamma$, where $\Psi_4$ will be
calculated in the two tetrads shown in Eq.~(\ref{eqn:bonditetradnull}) 
and (\ref{eqn:bonditetradnull2}).
We first start with the tetrad $\mathcal{T}_1$:
Figg.~(\ref{fig:newsbondi}) and (\ref{fig:newsdiff})
verify numerically the equivalence expressed by Eq.~(\ref{eqn:psi4gamma}):
it is clear that the two functions $\Psi_4$ and $-\gamma_{,vv}$ are
different in the non-linear regime but converge in the linear regime.
In particular Fig.~(\ref{fig:newsdiff}) shows in logarithmic scale
the absolute value of their difference at time $v=80$, well in the
linear regime. This numerical result proves the generic assumption
that $\Psi_4$ is related to the outgoing gravitational radiation
contribution.

\begin{figure}
\centering
\includegraphics*[width=8cm,height=7cm]{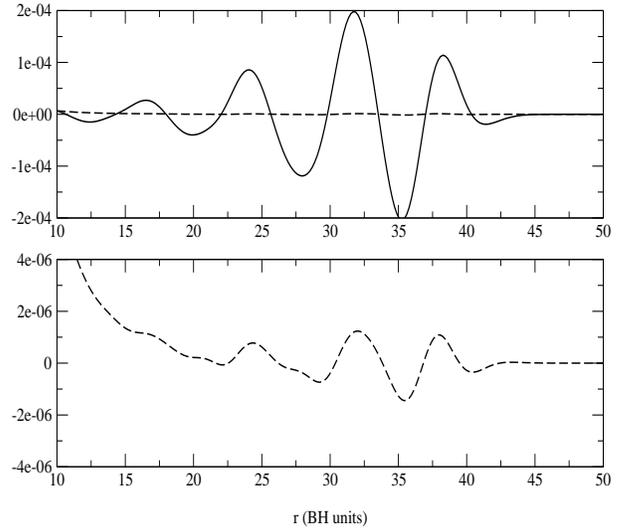}
\caption{(a) Top panel: the comparison of $\Psi_4$ (dashed line) 
calculated in tetrad
$\mathcal{T}_2$ and $-\gamma_{,vv}$ (solid line) for $v=80$. 
(b) Bottom panel: the value
for $\Psi_4$ alone.}
\label{fig:newsbondi2}
\end{figure}

As a counterexample, we show the same result when $\Psi_4$ is computed
in tetrad $\mathcal{T}_2$, which would actually have been our simplest
choice hadn't we applied the concept of a {\it quasi-Kinnesley tetrad}.
The results for this calculation are shown in Fig.~(\ref{fig:newsbondi2}).
It is evident that $\Psi_4$ does not get any contribution from the
background, meaning that the tetrad we have chosen is part of the
{\it quasi-Kinnersley frame}, however, it is evident that the result
is rather different from that coming from the Bondi function $\gamma$.

In order to understand what is happening, we need to analyze further
the tetrad $\mathcal{T}_2$, and in particular its limit when the
space-time approaches a type D one. Using our well known asymptotic
limits for the Bondi functions, it is easy to show that tetrad $\mathcal{T}_2$
converges, in the type D limit, to the tetrad

\begin{subequations}
\label{eqn:bonditetradnull2D}
\begin{eqnarray}
\ell^{\mu}&=&\left[
0,-1,0,0\right],
\label{eqn:bonditetradnulll2D} \\
n^{\mu}&=&\left[1,\frac{r-2M}{2r},
0,0\right],
\label{eqn:bonditetradnulln2D} \\
m^{\mu}&=&\left[0,0,\frac{1}{\sqrt{2}r}
,\frac{i}
{\sqrt{2}r\sin\theta}\right],
\label{eqn:bonditetradnullm2D}
\end{eqnarray}
\end{subequations}
which is different from the  Kinnersly tetrad used by Teukolsky, Eq.\ (\ref{eqn:schwarztetradnull}). A simple analysis
of the differences let us conclude that the original tetrad Eq.\ (\ref{eqn:schwarztetradnull}) can be obtained by first of all exchanging the two real
null vectors $\ell$ and $n$, and then using a boost transformation
of the type

\begin{subequations}
\label{eqn:boost}
\begin{eqnarray}
\ell&\rightarrow& A\ell, \label{eqn:boostl} \\
n&\rightarrow& A^{-1}n, \label{eqn:boostn}
\end{eqnarray}
\end{subequations}
where $A$ is a real parameter. It is easy to show that choosing 
$A=\frac{2r}{r-2M}$ we get that the new real null vectors coincide
with the Kinnersley tetrad defined in Eq.~(\ref{eqn:schwarztetradnull}). 
It is now straightforward to understand how these
differences affect the values of the Weyl scalars. First of all,
exchanging $\ell$ and $n$ corresponds to exchanging $\Psi_0$ and $\Psi_4$;
this means that if we use the tetrad $\mathcal{T}_2$ we will find the
outgoing radiative contribution in $\Psi_0$. This completely clarifies
the result found in Fig.~(\ref{fig:newsbondi2}): it turns out that
in this particular tetrad $\Psi_4$ is supposed to have a $r^{-5}$ radial
fall-off and, in practice, just like the result shown in 
Fig.~(\ref{fig:convpsi04}a) for $\Psi_0$ in tetrad $\mathcal{T}_1$, we are not
able to obtain this radial behaviour numerically, 
and we end up getting just numerical error. 

\begin{figure}
\centering
\includegraphics*[width=8cm,height=9cm]{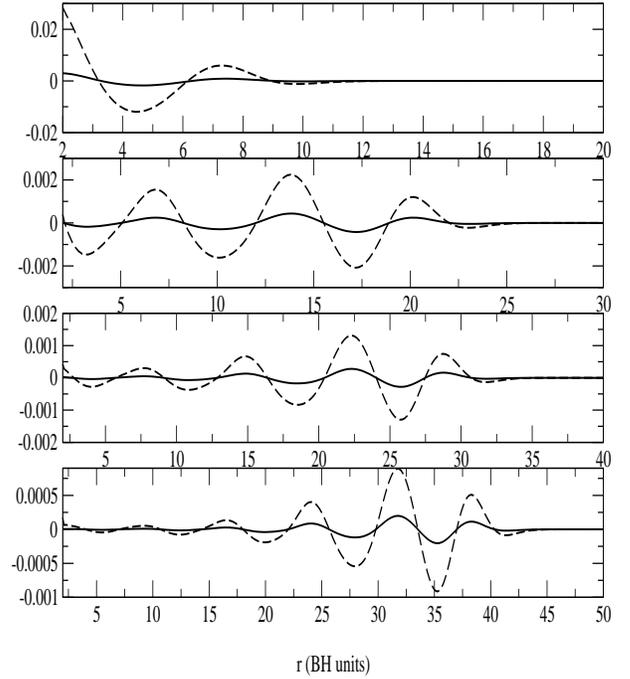}
\caption{From top to bottom: the comparison of $\Psi_0$ computed in the tetrad
$\mathcal{T}_2$ and $-\gamma_{,vv}$ for 
the time values of $v_0 = 10$, $v_1=40$, $v_2=60$ and $v_3=80$.
The dashed line is $\Psi_0$ while the solid line is $-\gamma_{,vv}$.}
\label{fig:newsbondi3}
\end{figure}

In Fig.~(\ref{fig:newsbondi3}) we show the comparison of $\Psi_0$ with
the news function; here the results are in better agreement 
but we still have no
correspondence, the reason for this is  to be found in 
the boost transformation, in fact a transformation like the one
written in Eq.~(\ref{eqn:boost}) changes the value of $\Psi_0$ 
according to

\begin{equation}
\Psi_0\rightarrow A^{-2}\Psi_0.
\label{eqn:psi0boost}
\end{equation}
This leads us to the final conclusion that, in the linearized regime,
the following relation must hold:

\begin{equation}
\left(\Psi_4\right)_{\mathcal{T}_1} = 
\left(\frac{r-2M}{2r}\right)^2\left(\Psi_0\right)_{\mathcal{T}_2} =
-\frac{\partial^2\gamma}{\partial v^2} 
\label{eqn:finalrel}
\end{equation}

\begin{figure}
\centering
\includegraphics*[width=8cm,height=9cm]{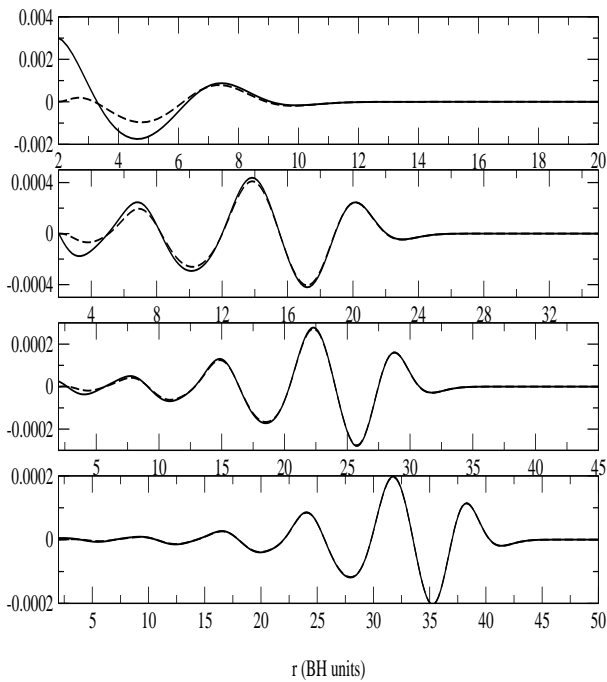}
\caption{From top to bottom: 
the comparison of $A^{-2}\Psi_0$ (dashed line) calculated in tetrad
$\mathcal{T}_2$ and $-\gamma_{,vv}$ (solid line) for
the values of $v_0 = 10$, $v_1=40$, $v_2=60$ and $v_3=80$.}
\label{fig:newsbondi4}
\end{figure}

We test this conclusion in  Fig.~(\ref{fig:newsbondi4}) where we have
plotted the value of 
$\left(\frac{r-2M}{2r}\right)^2\left(\Psi_0\right)_{\mathcal{T}_2}$.

We emphasize that all the results that we
have obtained in all the tetrads have wave-like profiles, although
only one is the correct wave contribution. In practical numerical
simulations one should really make sure that the tetrad in which
the scalars are computed is a {\it quasi-Kinnersley} tetrad,
otherwise the results, even if wave-like shaped, could be wrong.

\subsection{Energy calculation}
\label{sec:energy}

Having made sure that $\Psi_4$ calculated in tetrad 
$\mathcal{T}_1$ is related, in the linear regime, to the
Bondi news function, we can use its expression to calculate
the energy radiated from the black hole. In section \ref{sec:psinews}
we have shown that the expression of the energy flux per unit solid
angle is given by

\begin{equation}
\frac{d^2E}{dvd\Omega} = -\frac{r^2\Phi^2}{4\pi},
\label{eqn:energyfluxgen}
\end{equation}
where we denote with $\Phi$ the generic expression for the news function,
being it $\gamma_{,v}$ or $\int\left(\Psi_4\right)_{qKT}dv$. We can integrate
the expression in Eq.~(\ref{eqn:energyfluxgen}) on a 2-sphere in order to
obtain the energy flux. For the sake of simplicity we take a sphere
of radius $r_0$, getting the result

\begin{equation}
\frac{dE}{dv} = -\frac{r_0^2}{4\pi}\oint\Phi^2\sin\theta d\theta d\phi,
\label{eqn:energyfluxgen2}
\end{equation}
 
The computation of the energy flux  was the goal of \cite{Papadopoulos02}, where the news function was used to 
calculate the amount of energy which is carried away by each
spin-weighted spherical harmonics of the outgoing radiation.
We can perform a similar calculation using $\Psi_4$ and
compare our results, in order to have a further demonstration
of the validity of our approach. 
Given the results described in Section \ref{sec:psi4news}
it is evident that also these results will be in good agreement,
however we want to highlight their validity and to show their
dependence on the position of the observer.

\begin{figure}
\centering
\includegraphics*[width=8cm, height=8cm]{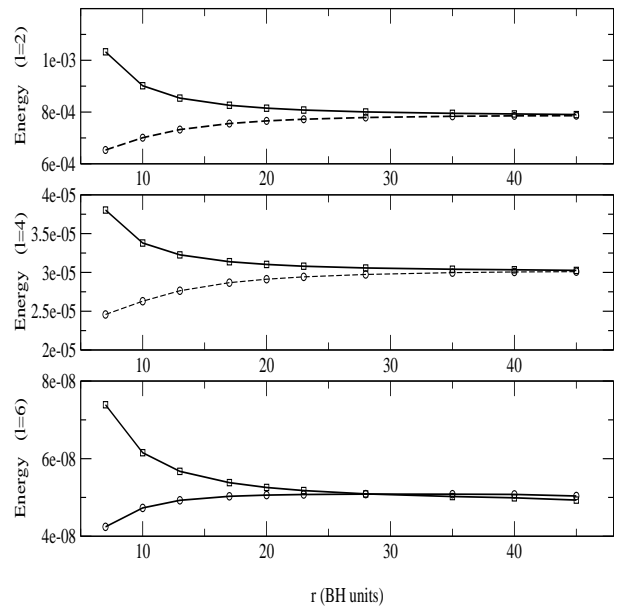}
\caption{From top to bottom: energy contribution of the $l=2$ 
harmonic initial data. 
The three graphs represent the outgoing energy contribution for the
values of $l=2,4,6$. In each graph the two curves represent  
the value for the energy calculated using the
Bondi news function (upper curve), while the lower curve uses
the value of $\Psi_4$ in tetrad $\mathcal{T}_1$, as a function
of the position of the observer. It is evident that at late times
there is a convergence of the two values. This convergence seems to
be less evident for the $l=6$ case (lowest graph), but we expect this
phenomenon to be purely numerical, because the numerical error on this
multipole component is very high.}
\label{fig:energyl2T1}
\end{figure}

\begin{figure}
\centering
\includegraphics*[width=8cm, height=6cm]{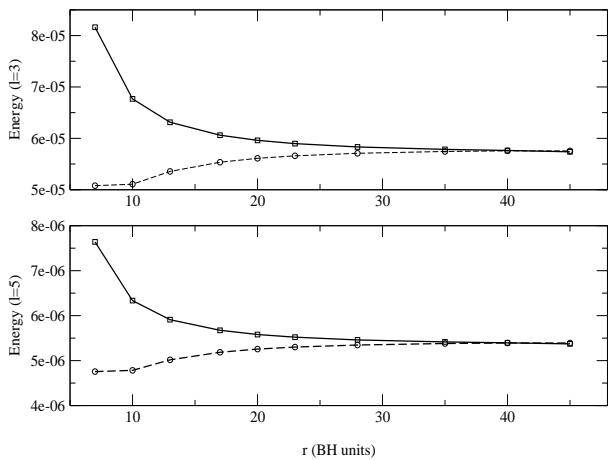}
\caption{From top to bottom: energy contribution 
of the $l=3$ harmonic initial data.
We show the two graphs corresponding to the dominant terms $l=3,5$
in the emitted gravitational signal. Again here we compare the 
result coming from the Bondi news function with the one using
$\Psi_4$ in tetrad $\mathcal{T}_1$. The results are similar to the
ones shown in Fig.~(\ref{fig:energyl2T1}).}
\label{fig:energyl3T1}
\end{figure}

Since we are interested in the energy contribution of each spin
weighted spherical harmonic, we first have to perform the decomposition
of the signal into spin weighted spherical harmonics contributions.
This is done by introducing the quantity $\Phi_l$ defined as

\begin{equation}
\Phi_l\left(v,r\right)=2\pi\int^1_{-1}
\Phi\left(v,r,y\right)Y_{2l0}\left(y\right)dy,
\label{eqn:spinangularmode}
\end{equation}
where $y=-\cos\theta$ and $Y_{2l0}$ is the spin weighted spherical
harmonic of spin 2.

Using 
Eqq.~(\ref{eqn:energyfluxgen2}) and (\ref{eqn:spinangularmode}) we can get 
an expression for the
total energy emitted in each angular mode after the evolution
to a final time $T$, given by

\begin{equation}
E_l\left(T\right)=\frac{r_0}{4\pi}\int^T_0
\left[\Phi_l\left(v,r=r_0\right)\right]^2dv.
\label{eqn:enangularmode}
\end{equation}
We have performed some numerical simulations where the energy, using
both the Bondi news function and $\Psi_4$, has been calculated.
The results are shown in Figg.~(\ref{fig:energyl2T1}) and 
(\ref{fig:energyl3T1}).

Fig.~(\ref{fig:energyl2T1}) shows the result for a numerical
simulation where the initial profile of the $\gamma$ has been 
chosen to be quadrupolar, i.e. using the spin-weighted spherical
harmonic with $l=2$, $m=0$. The non-linearity of the problem
is translated into the fact that the evolution excites higher
order multipolar terms. However, simmetry considerations allow
only even multipolar terms to be excited. In the picture we show
the energy at time $v=80$ for the $l=2,4,6$ terms. Such energy 
is calculated varying the position of the observer and it is
evident that, as soon as we push the observer further from the
source, the two energy calculations coincide. On the other hand,
numerical errors become stronger when going higher order multipole
terms, which explains the not-perfect convergence for the $l=6$
terms.

Fig.~(\ref{fig:energyl3T1}) shows a similar simulation
for an initial data with $l=3$, $m=0$. Here again we expect the
non-linearity to excite the other harmonics. Differently from the 
$l=2$ case, we don't expect to have forbidden modes, however our
numerical results show that the highest amplitude modes are 
the odd ones, so we show only those modes. Anyway the results
in this case are qualitatively equivalent to those obtained
in the case of quadrupolar initial data.


\section{Conclusions}
\label{sec:concl}

The problem of correctly extracting the gravitational 
signal in numerical simulations
is of primary importance. We believe that the use of the  Weyl scalars of the Newman-Penrose formalism
offers a promising method for wave extraction,  as it applies to any formulation of Einstein's
equations and, even more important, to any kind of background we
end up with, being it Schwarzschild or Kerr. However the
problem of identifying the tetrad in which one is to compute the Weyl scalars 
still awaits a full solution. Recent work \cite{Beetle04, Nerozzi04}
shows how to make the important first step of  identifying an equivalence class of tetrads,
the quasi-Kinnersley frame, of which the desired quasi-Kinnersley tetrad is a member.  However,
the problem of isolating the right tetrad out of this set is still under investigation.

In the present work we have considered a non-trivial numerical scenario, namely
the evolution of a non-linearly perturbed black hole using Bondi
coordinates, in order to show  the importance
of the tetrad choice for the calculation of wave related quantities.
This particular scenario is well suited for a practical demonstration of
the problems one  would encounter  if a  careful choice of the tetrad
for the Weyl scalar computation is not  done. We have in fact shown
that the computation of the Weyl scalars in an arbitrary  tetrad, chosen by brute-force  using 
mathematical packages like GrTensor, would lead to wrong results for
$\Psi_4$, which is the quantity that typically is  supposed to contain the (outgoing) 
gravitational wave degrees of freedom. 
This fact is evident in our case, where we have compared directly the
results for the Weyl scalar in two different tetrads $\mathcal{T}_1$ and $\mathcal{T}_2$, 
using the Bondi news function in determining that the Weyl scalar corresponding to the quasi-Kinnersley tetrad is the right one. In Section \ref{sec:results} we have shown that only the 
quasi-Kinnersley tetrad $\mathcal{T}_1$ gives results in agreement with the news function.

Finally, we emphasize the importance of singling out  an appropriate 
quasi-Kinnersley tetrad from the quasi-Kinnersley frame \cite{Beetle04,
Nerozzi04}. For instance, the tetrad $\mathcal{T}_2$ after exchange of the $\ell$ and $n$ null basis vectors is related to the tetrad $\mathcal{T}_1$ by a boost. This example indicates that  every
tetrad in the quasi-Kinnersley frame will give  results for $\Psi_0$
and $\Psi_4$ that will show no contribution from the background, so that
the wave-like shape of the scalars could lead us to the wrong conclusion
of having the right outgoing gravitational signal.  As we have shown, this conclusion could well be far from reality.


\acknowledgments

The authors are indebted to Philippos Papadopoulos and Denis Pollney for their help
and invaluable discussions.
AN is 
funded by the NASA grant NNG04GL37G to the University of Texas
at Austin and by the EU Network Programme (Research Training Network
contract HPRN-CT-2000-00137). MB is partly funded by 
MIUR (Italy). Work on this research started when LMB was at Bates College.


\bibliographystyle{apsrev}
\bibliography{references}


\end{document}